\documentclass[reprint, aps, amsmath, amssymb, showpacs]{revtex4-1}

\usepackage{amsmath,amssymb,graphicx,lipsum}
\usepackage{amsbsy}
\usepackage{latexsym}
\usepackage{xcolor}
\usepackage{graphicx}
\usepackage{psfrag}
\usepackage[normalem]{ulem}
\usepackage{bm}
\usepackage{hyperref, dblfloatfix}
\usepackage{float}

\begin{document}
\title{Accelerated Inertial Regime in the Spinodal Decomposition of Magnetic Fluids}  
\author{Anuj Kumar Singh}
\author{Varsha Banerjee}
\affiliation{Department of Physics, Indian Institute of Technology Delhi, New Delhi, 110016 India}

\begin{abstract}

Furukawa predicted that at late times, the domain growth in binary fluids scales as $\ell(t)\sim t^{2/3}$, and the growth is driven by fluid inertia. The {\it inertial growth regime} has been highly elusive in molecular dynamics (MD) simulations. We perform coarsening studies of the ($d=3$) Stockmayer (SM) model comprising of magnetic dipoles that interact via long-range dipolar interactions as well as the usual Lennard-Jones (LJ) potential. This fascinating polar fluid exhibits a gas-liquid phase coexistence, and magnetic order even in the absence of an external field. From comprehensive MD simulations, we observe the inertial scaling [$\ell(t)\sim t^{2/3}$] in the SM fluid for an extended time window. Intriguingly, the fluid inertia is overwhelming from the outset - our simulations do not show the early diffusive regime [$\ell(t)\sim t^{1/3}$] and the intermediate viscous regime [$\ell(t)\sim t$] prevalent in LJ fluids.   
\end{abstract}
\maketitle

\section{introduction}
When quenched below the spinodal temperature, a homogeneous fluid separates into a low-density gas phase that coexists with a high-density liquid phase. The system evolves towards a new equilibrium state, and this evolution involves capillary forces, viscous dissipation, and fluid inertia.  The co-existing phases or domains grow with time and form bi-continuous structures with sharp well-defined interfaces. If the domain morphology remains unchanged with time, the system exhibits dynamical scaling \cite{binder1974}. The growth process is then described by a unique length scale $\ell(t)$. It typically grows as a power-law: $\ell(t)\sim t^{\alpha}$, where the exponent $\alpha$ depends on the transport mechanism that is dominant during the coarsening (or segregation) process. The late-time behavior of this generic system undergoing spinodal decomposition continues to have open questions despite many theoretical, experimental, and computational studies \cite{siggia1979, furukawa1985, furukawa1987, onuki2002}. 

Diffusive growth in phase-separating solid mixtures is captured by the Lifshitz-Slyozov (LS) law \cite{lifshitz1961}: $\ell(t)\sim t^{1/3}$. In fluids and polymers however, hydrodynamic effects become important after the initial diffusive regime. It was argued by Furukawa that fluid inertia is negligible compared to the fluid viscosity for early times, while the reverse is true for late times \cite{furukawa1985}. A dimensional analysis leads to the following additional growth regimes: $\ell(t)\sim t$ for $\ell(t)\ll \ell_{i}^*$;  $\ell(t)\sim t^{2/3}$ $\ell(t)\gg \ell_{i}^*$. The inertial length scale $\ell_{i}^*=\eta^2/\Tilde{\sigma}\rho$, where  $\Tilde{\sigma}$ is the interfacial tension, $\rho$ is the fluid density and $\eta$ is the shear viscosity. It marks the cross-over from a low-Reynolds number ($R=\rho/\eta\ell$) viscous hydrodynamic regime to an inertial regime \cite{grant1999}. 

An outstanding issue in the spinodal decomposition of bulk fluids is the evidence of the theoretically predicted inertial growth in experiments and computations. Experimentally, it has been reported in dewetting kinetics of polymer thin films \cite{Lal2020, Reiter2001} and domain coarsening during spinodal decomposition in binary fluid mixtures \cite{Malik1998, Livet2001}. Theoretically, the first observation of linear growth in the viscous regime was in the numerical studies of the phenomenological {\it Model H} \cite{puri1992}. On the other hand, both viscous and inertial regimes were observed in lattice Boltzmann simulations which augment the Cahn-Hillard equation with the Navier-Stokes equations to model the velocity field \cite{Kendon1999, Kendon2001}. Molecular dynamics (MD), which includes details of microscopic physics in following the motion of each particle and in-built hydrodynamics, has been used much less to study domain growth due to the heavy computational requirements. MD simulations of Lennard-Jones-like binary fluids ($d=2$) reported inertial growth at late times for critical quenches \cite{velasco1993,velasco1996}. However the evaluations of the exponent is not conclusive, and the 2/3 law did not survive an averaging process over independent realizations \cite{ossadnik1994}. Few studies ($d=3$) have reported linear viscous growth \cite{laradji1996,ahmad2010,ahmad2012,majumder2011}, but the much sought-after inertial regime has remained elusive.   

In this article, we present comprehensive MD simulations to study the spinodal decomposition in a ($d=3$) polar fluid comprising of magnetic dipoles, namely the Stockmayer (SM) fluid. The dual properties of being fluid and having magnetic order even in the absence of external fields make it an intriguing system \cite{Stevens1995}. The SM fluid is realized by ferrofluids and other magneto-rheological fluids which have applications and technological promise.  The SM particles experience short-range isotropic attractive interactions and long-range dipole-dipole anisotropic interactions. Detailed investigations have revealed that for sufficient concentration of the magnetic dipoles, the SM fluid undergoes a gas-liquid (GL) phase transition on cooling \cite{Stevens1995, Leeuwen1993, samin2013, Bartke2007}. So what are the consequences of the long-range dipole-dipole interactions on the coarsening magnetic liquid phase? We initiate this inquiry by quenching the paramagnetic gas $(T>T_c)$ into the coexistence region $(T<T_c)$. Our novel observations are: (i) The coarsening morphologies exhibit an {\it accelerated inertial growth law} $\ell_s(t)\sim t^{2/3}$ for nearly two decades, hitherto unobserved in MD simulations; (ii) {\it Triggered magnetic order} in the liquid phase that grows as $\ell_M(t)\sim t$, typical of dipolar magnets with non-conserved order parameter dynamics \cite{bray1994, bupathy2017}. In what follows, we will focus on understanding these fascinating observations. Our paper is organized as follows. Sec.~2 provides the model and the simulation details. The detailed numerical results are provided in Sec.~3. Finally, Sec.~4 contains the summary of results and the conclusion.

\section{Model and Simulation Details}
Let us consider a collection of $N$ magnetic dipoles with mass $m$ and magnetic moment $\vec{\mu}=\mu\hat{\mu}$. In the SM model, the interaction potential between particles $i$ and $j$ separated by $\vec{r}_{ij}=r_{ij}\hat{r}_{ij}$ is represented by \cite{Leeuwen1993}:
\begin{align}
\label{SM}
U(\vec{r_{ij}}, \vec{\mu_i}, \vec{\mu_j}) &= 4\epsilon\sum_{i,j}\bigg[{\bigg( \frac{\sigma}{r_{ij}}\bigg)}^{12}-{\bigg( \frac{\sigma}{r_{ij}}\bigg)}^6\bigg] \\ \nonumber 
&+\frac{\mu_0\mu^2}{4\pi}\sum_{i,j}\bigg[ \frac{\hat{\mu}_i.\hat{\mu}_j-3(\hat{\mu}_i.\hat{r}_{ij})(\hat{\mu}_j.\hat{r}_{ij})}{r_{ij}^3} \bigg].
\end{align}
The first two terms describe the usual Lennard-Jones (LJ) potential energy comprising of the short-range steric repulsion and weak van der Waals attraction. The parameters $\sigma$ (particle diameter) and $\epsilon$ (depth of the attractive potential) set the units of length and energy in our study. The third term  represents the dipole-dipole interactions which are significant up to large distances, and can be 0, $\pm$ depending on the position and orientation of the dipoles $i$ and $j$. The particles thus experience isotropic short-range van der Waal's attraction as well as anisotropic long-range dipolar interactions. When cooled below the critical temperature $T_c$, the SM fluid undergoes a phase transition from a paramagnetic gas phase to a GL co-existence phase. This phase diagram in the $\rho-T$ plane has been determined for a range of $\mu$ values using Monte Carlo and MD simulations  \cite{watanabe2012, Stevens1995, Leeuwen1993, samin2013, Bartke2007}. The primary effect of increasing $\mu$ is to shift the critical point $(\rho_c,T_c)$ upwards, thereby enlarging the GL co-existence region. As will be discussed in Sec.~3, the qualitative behaviour is not affected by the strength of $\mu$.

We have performed large-scale MD simulation of the SM fluid ($d=3$) in the canonical $(NVT)$ ensemble using LAMMPS \cite{LAMMPS}. The simplest Langevin thermostat does not incorporate hydrodynamics. Popular for the calculation of transport properties have been the Nos\'e-Hoover thermostat (NHT) and dissipative particle dynamics (DPD) \cite{binder1996, frenkel2001,binder2004}. The NHT is not Galilean-invariant and hence conserves only the total momentum. However, it exhibits excellent temperature control crucial for constant temperature ensembles. DPD on the other hand is Galilean-invariant and preserves is the local momentum, but has problems with maintaining constant temperature. Comparative studies of the two thermostats have revealed that even at criticality, diffusivity and shear viscosity show excellent agreement \cite{roy2015}. So inspite its problems and newer protocols \cite{allen2007}, the NHT continues to remain popular for the study of domain growth. The coarsening phenomenon occurs at large length scales and time scales and is not critically dependent on the microscopically exact replication of hydrodynamics. As a result, several MD simulations of the LJ fluid using the NHT have correctly reproduced the theoretical predictions of the early diffusive growth [$\ell_s(t)\sim t^{1/3}$] followed by the viscous regime [$\ell_s(t)\sim t$] \cite{ahmad2010,ahmad2012,majumder2011}. 

Our simulations have been performed using the NHT. The dipolar sums in Eq.~(\ref{SM}) have been computed using the Ewald summation technique with metallic boundaries \cite{frenkel2001}. We have taken a cubic  box with periodic boundary conditions of volume $V = 75^3$ (in LJ units) with $N$ = 84375, 126563, 168750 SM particles corresponding to density $\rho$ = 0.2, 0.3, 0.4 respectively. These values of density fall in the spinodal region. The calculations are performed in reduced units defined as: $ T^*= k_BT/\epsilon$, $ \rho^*= N\sigma^3/V $, $ \mu^*= \mu/\sqrt{\epsilon \sigma^3} $,  $ \Delta t^*= \Delta t/\sqrt{m\sigma^3/\epsilon} $. (We drop the star in the subsequent discussions.) The MD runs were performed using the standard velocity-Verlet algorithm with simulation time step $\Delta t = 0.002$ \cite{velocity-verlet1983}. 
We present results for a prototypical value of $\mu=2.5$ and use the GL co-existence data from Ref. \cite{Stevens1995} which reports the critical point as $\rho_c=0.29(1)$, $T_c=2.63(1)$. 
Starting with a random orientation of particles, the system is first equilibrated at a high temperature ($T=5$) to obtain an isotropic and homogeneous initial state. It is then quenched into the GL coexistence regime ($T<T_c$) and evolved up to $t=500$ (or $2.5 \times 10^5$ steps) to observe the coarsening. (Finite size effects set in soon after.) All data has been averaged over 5 independent runs.

\section{Detailed Numerical Results}
\begin{center}
\begin{figure}
    \centering
        \includegraphics[width=0.8\linewidth]{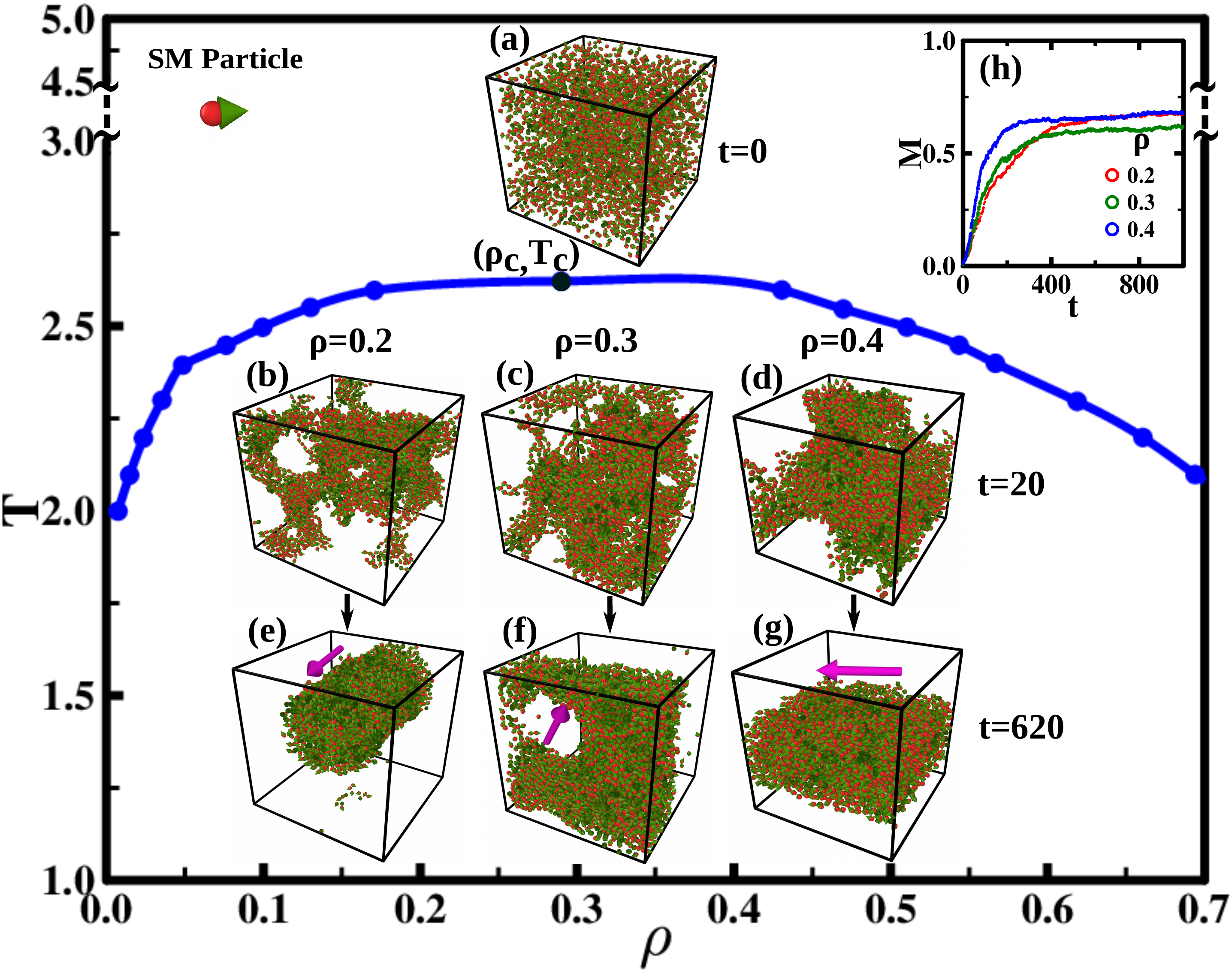}
    \caption{Gas-liquid (GL) coexistence region of SM fluid for dipole moment $\mu=2.5 $ from Ref. \cite{Stevens1995} in reduced LJ units. (a) Typical initial state at $T=5$. (b)-(d) Early time ($t=20$) coarsening morphologies for indicated values of $\rho$. (e)-(g) are corresponding equilibrated structures ($t=620$). (h) $\mathbf{M}$ vs. $t$ indicating the growth of magnetic order. Magenta arrows provide the direction of magnetization $\hat{\mathbf{M}}$ in the equilibrated morphologies.}
    \label{1}
\end{figure}
\end{center}

It is useful to know the equilibrium morphologies before addressing the non-equilibrium evolution. Fig.~\ref{1} shows the $\mu=2.5$  coexistence curve with data read from Ref. \cite{Stevens1995}. A representative initial configuration ($t=0$) corresponding to the paramagnetic gas ($T>T_c$) is shown in sub-figure (a). (Smaller cubic box with $L=24$ has been used for clarity in visualization.)  Spinodal decomposition is initiated by a deep quench to $T=1.05$. The coarsening morphologies at early time ($t=20$) for $\rho$ = 0.2, 0.3 and 0.4 are shown in the sub-figures (b)-(d). They exhibit bi-continuous structure that is characteristic of spinodal decomposition. When evolved for long ($t=620$), distinct equilibrated structures are obtained for different densities: (e) cylindrical for $\rho=0.2$, (f) inter-penetrating cylinders for $\rho=0.3$ and (g) planar for $\rho=0.4$. Some of these density-dependent shapes have been observed in earlier studies of the LJ fluid \cite{das2002, macdowell2006, schrader2009, block2010, majumder2010, binder2012, roy2013} as well as the SM fluid \cite{richardi2009, salzemann2009}. Another significant observation is the development of magnetization ${\bf M}=\sum_{i=1}^N \vec{\mu}_i/N$ with time, see $M(t)$ vs. $t$ behaviour in the inset (h). The magenta arrows in the equilibrated morphologies indicate the unit vector $\hat{\mathbf{M}}$ at equilibrium. 

\begin{figure}
\begin{center}
 \includegraphics[width=1\linewidth]{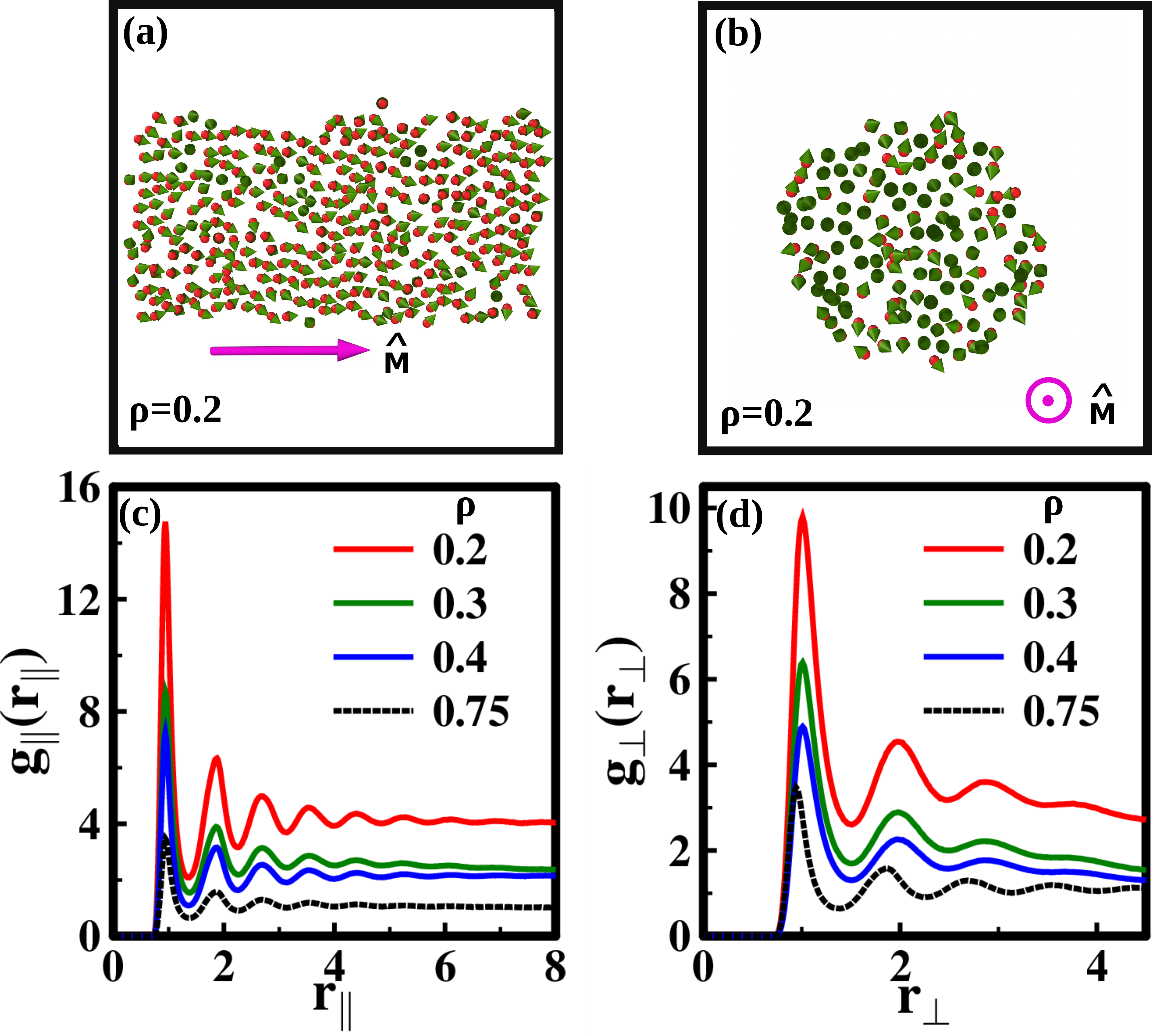}
 \caption{Typical morphology slices (a) parallel and (b) perpendicular to $\hat{\mathbf{M}}$. (c) $g_{\parallel}(r_{\parallel})$ vs. $r_{\parallel}$ and (d) $g_{\perp}(r_{\perp})$ vs. $r_{\perp}$ for the equilibrated morphologies in Figs.~\ref{1}(e)-1(g). The dashed lines for $\rho=0.75$ indicate that $g_{\parallel}(r_{\parallel})$ and $g_{\perp}(r_{\perp})$ approach 1 for larger densities, see related text for details.}
\label{2}
\end{center}
\end{figure}

Let us understand the spatial and magnetic order in the anisotropic morphologies obtained at equilibrium. In Figs.~\ref{2}(a) and \ref{2}(b) we show a typical slices taken parallel and perpendicular to $\hat{\mathbf{M}}$ for the cylindrical morphology. The chain-like alignment of dipoles along $\hat{\mathbf{M}}$ is unmistakable. (These observations are also borne by the other morphologies.) The standard probe to confirm the formation of the liquid state is the pair correlation function \cite{Weis1993}:

\onecolumngrid
\twocolumngrid

\begin{equation}
\label{PCF}
g(r)=\frac{1}{N\rho_0}\bigg\langle\sum_{\stackrel{i,j}{i\neq j}}^N\frac{\delta(r-r_{ij})}{(4/3)\pi[(r+\Delta r)^3-r^3]}\bigg\rangle.
\end{equation}
The $\delta$ function is unity if $r_{ij}$ falls within the shell of thickness $\Delta r$ centered at $r$ and zero otherwise.  The division by $N$ ensures that $g(r)$ is normalized to a per particle function. By construction, $g(r)=1$ for an ideal gas, and any deviation implies correlations between the particles due to the inter-particle interactions. In the liquid phase, $g(r)$ exhibits a large peak at small-$r$ signifying nearest neighbour correlations followed by small oscillations which eventually approach 1 at large-$r$ \cite{Weis1993}. Defining $r_{\parallel}$ and $r_{\perp}$ to be distances along and perpendicular to $\hat{\mathbf{M}}$, we show $g_{\parallel}(r_{\parallel})$ vs. $r_{\parallel}$ in Fig.~\ref{2}(c) and $g_{\perp}(r_{\perp})$ vs. $r_{\perp}$ in Fig.~\ref{2}(d). These evaluations indicate that the aggregates are in the liquid phase. Further, the presence of peaks at multiples of the particle diameter reconfirm the formation of chains along $\hat{\mathbf{M}}$ for all the morphologies. Note that $g(r)$ in our evaluations does not approach 1 as our equilibrated structures are small. This is not the case for larger densities, e.g. $\rho = 0.75$, for which the equilibrated structure fills the box. The corresponding evaluations are shown by dashed lines in Figs.~\ref{2}(c)-(d).

A standard probe to characterize coarsening morphologies is the  two-point equal-time correlation function \cite{bray2002,puri2009}: 
\begin{align}
C(\boldsymbol{r}_i, \boldsymbol{r}_j, t)=\langle \psi(\boldsymbol{r}_i).\psi(\boldsymbol{r}_j)\rangle - \langle\psi(\boldsymbol{r}_i)\rangle \langle\psi(\boldsymbol{r}_j)\rangle,
\end{align}
where $\psi(\boldsymbol{r})$ is the appropriate order parameter, $\boldsymbol{r}=\boldsymbol{r}_i-\boldsymbol{r}_j$ and the angular brackets indicate an ensemble average. The characteristic length scale $\ell(t)$ is defined from the correlation function as the distance over which it decays to a suitably chosen fraction of its maximum value. If the correlation function obeys dynamical scaling, $C(r,t) = f\left(r/\ell\right)$, where $f(x)$ is a scaling function \cite{bray2002,puri2009}. The coarsening morphologies will be scale-invariant and characterized by the unique length scale $\ell(t)$. An equivalent probe is the structure factor $S(\textbf{k},t)$, which is the Fourier transform of $C({\textbf{r}},t)$. The corresponding dynamical-scaling form is $S(k,t) = \ell^{d}\tilde{f}\left(k\ell\right)$, where $\tilde{f}(p)$ is the Fourier transform of $f(x)$. For a scalar order parameter, $S(k) \sim k^{-(d+1)} \quad \mbox{as} \quad k \rightarrow \infty$. This result, referred to as the {\it Porod law}, signifies scattering off sharp interfaces \cite{bray2002, puri2009}. 

\begin{figure}
    \begin{center}
    \includegraphics[width=1\linewidth]{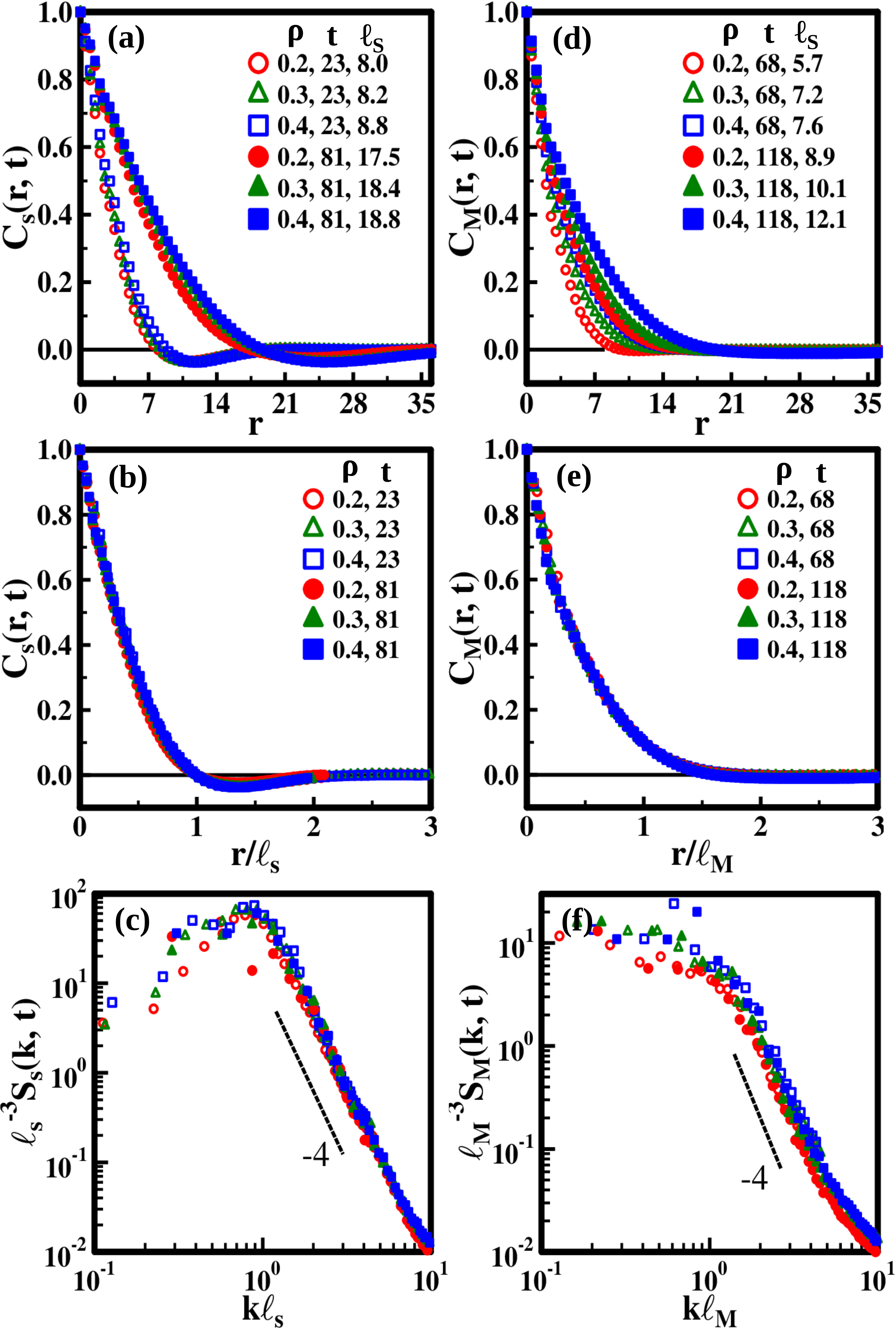}
    \caption{(a) Unscaled spatial correlation function $C_s(r, t)$ vs. $r$, (b) corresponding scaled correlation function $C_s(r, t)$ vs. $r/\ell_s$ and (c) scaled structure factor $S_s(k,t)$ vs. $k\ell_s$ on a log-log scale. (d) Unscaled magnetic correlation function $C_M(r, t)$ vs. $r$, (e) corresponding scaled correlation function $C_M(r, t)$ vs. $r/\ell_M$ and (f) scaled structure factor $S_M(k,t)$ vs. $k\ell_M$. Dashed lines in the structure factor plots denote the Porod tails.}    
    \label{3}
\end{center}
\end{figure}

The equilibrated morphologies in Fig.~\ref{1} have spatial as well as magnetic order, so we evaluate the spatial correlation lengthscale $\ell_s$ and the magnetic correlation lengthscale $\ell_M$. For this purpose, the continuum system is mapped onto a spin-lattice by discretizing the volume $V$ into sub-boxes of size $2^3$. (Our results do not depend on the size of the sub-box.) A sub-box $i$ centered at $\boldsymbol{r}_i$ with density $\rho_i>\rho$ is identified as liquid phase with $\psi_s(\boldsymbol{r_i})=1$. On the other hand, $\rho_i<\rho$ is identified as the gas phase with $\psi_s(\boldsymbol{r}_i)=-1$. For the magnetic order in the liquid phase, the order parameter $\psi_{M}(r_i)$ is the average dipole moment of the particles in the sub-box $i$. Fig.~\ref{3}(a) shows the evaluation of the correlation function $C_s(r, t)$ vs. $r$ for specified values of $\rho$ and $t$. We have defined the average domain length for the liquid phase $\ell_s$ as the first zero crossing of the correlation function $C_s(r, t)$. This value for each data set has also been specified in the legend. The corresponding scaled correlation function $C_s(r, t)$ vs. $r/\ell_s$ is shown in Fig.~\ref{3}(b). The small dip in $C(r)$ is characteristic of periodic modulations in bi-continous morphologies \cite{bray2002,puri2009}.  The system exhibits dynamical scaling for all values of $\rho$ indicating the presence of a unique lengthscale. The data also scale for the different values of $\rho$.  The corresponding scaled structure factor $S_s(k,t)$ vs. $k\ell_s$ shown in Fig.~\ref{3}(c) has a Porod tail, $S_s(k)\sim k^{-4}$ due to scattering from smooth GL interfaces. 

Similarly, Fig.~\ref{3}(d) shows the corresponding magnetic correlation  $C_M(r, t)$ vs. $r$ for indicated values of $\rho$ and $t$. The average magnetic domain size $\ell_M$ is also provided. It is defined as 0.1 of the maximum value of correlation function $C_M(r, t)$. The scaled magnetic correlations  $C_M(r, t)$ vs. $r/\ell_M$ are shown in Fig.~\ref{3}(e). This data also exhibits dynamical scaling, and scales for different values of $\rho$.  Further, the corresponding  $S_M(k,t)$ vs. $k\ell_M$ shown in Fig.~\ref{3}(f) also exhibits a Porod tail $S(k)\sim k^{-4}$. It should be mentioned that for an $n$-component order parameter, the tail is expected to obey the generalized Porod law: $S(k)\sim k^{-d+n} \equiv k^{-6}$ characteristic of scattering from monopoles and hedgehogs \cite{bray2002, puri2009}. The morphologies obtained from our simulations have smooth GL interfaces. Consequently, the interfacial scattering $S_M(k)\sim k^{-4}$ dominates.  

\begin{figure}
    \begin{center}
    \includegraphics[width=0.65\linewidth]{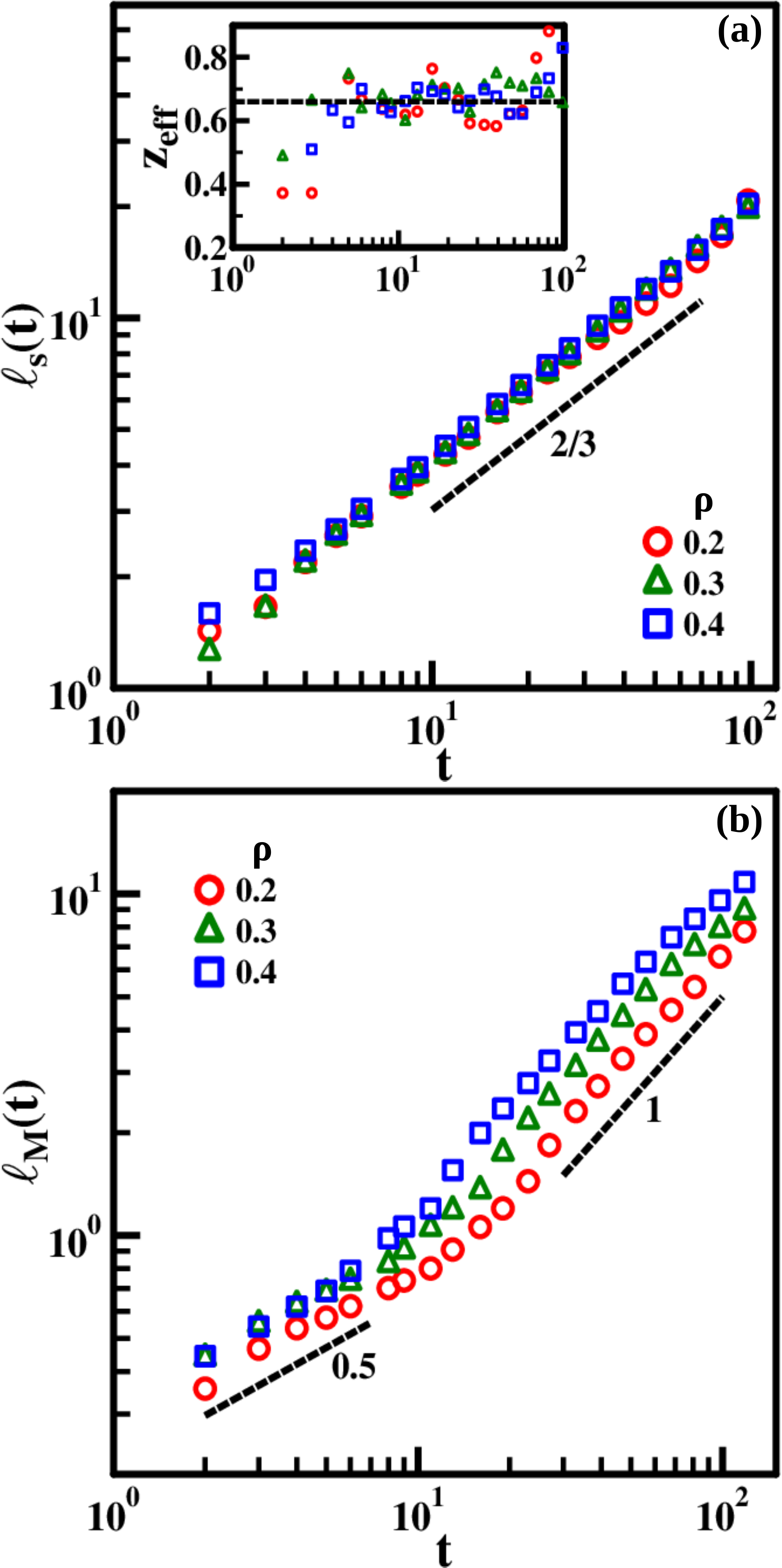}
    \caption{(a) Plot of $\ell_s(t)$ vs. $t$  on a log-log scale. The inset represents the effective growth exponents $z_{eff} =\partial \ln \ell_s/\partial \ln t$ vs. t on a log-linear scale. The dashed line with slope 2/3 corresponds to inertial growth. (b) Plot of $\ell_M(t)$ vs. $t$ on a log-log scale. The dashed lines with specified slopes are a guide to the eye, see text for details.}
    \label{4}
\end{center}
\end{figure}

We now present our most striking result. Fig.~\ref{4}(a) shows the growth of spatial correlations $\ell_s(t)$ vs. $t$ on a log-log scale for $\rho = 0.2, 0.3, 0.4$.  To accurately determine the growth law, we evaluate the effective growth exponent $ z_{\rm eff}=\partial \ln \ell_s/\partial \ln t$. This evaluation, shown in the inset, yields a value of $\bar{z}\simeq 2/3 (\simeq 0.66)$, also shown by the dashed line in the main figure. The data obey $\ell_s(t)\sim t^{2/3}$ for more than a decade. Though predicted by Furukawa in 1985 \cite{furukawa1985}, inertial growth law has not been observed in MD simulations  \footnote{In Ref. \cite{roy2013}, the authors speculated the $t^{2/3}$ inertial growth law in the coarsening LJ fluid for a value of $\rho = 0.16$ close to the spinodal line. This could not be unambiguously demonstrated in their simulations.}. Another significant feature is that the inertial growth is {\it accelerated}: the customary diffusive [$\ell_s(t)\sim t^{1/3}$] and viscous [$\ell_s(t)\sim t$] regimes are not observed in our simulations. Fig.~\ref{4}(b) shows the growth of magnetic correlations $\ell_M(t)$ vs. $t$ which are delayed as compared to spatial ordering. In a comprehensive study of growth laws for systems with long-range interactions \cite{bray1994}, dipolar solids with non-conserved dynamics were found to follow the growth law: $\ell(t)\sim t$. The dashed line with slope 1 is a guide to the eye. The limited data suggests that growth of (delayed) magnetic order $\ell_M(t)\sim t$, but larger system sizes will be required to observe this behaviour over an extended time window.

\begin{figure}
    \begin{center}
    \includegraphics[width=1\linewidth]{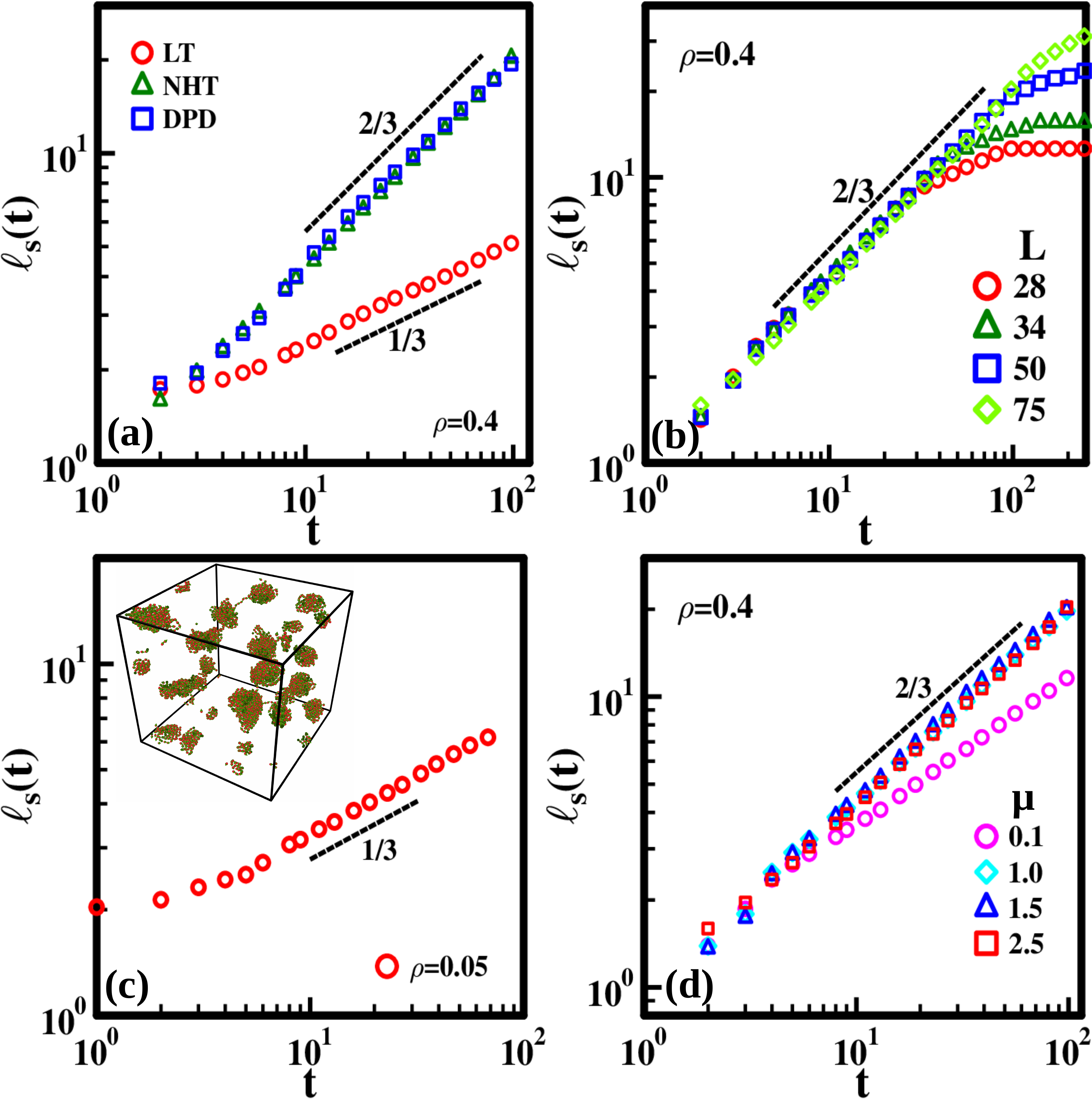}
    \caption{(a) Comparison of $\ell_s(t)$ vs. $t$ for Langevin thermostat (LT), Nos\'e Hoover thermostat (NHT), and dissipative particle dynamics (DPD). (b) Plot of $\ell_s(t)$ vs. $t$  for different values of $L$. (c) Length scale $\ell_s(t)$ vs. $t$ in nucleation and growth regime. The inset shows a typical intermediate time ($t=60$) morphology in this regime. (d) Plot of $\ell_s(t)$ vs. $t$ for dipole moments $\mu$=0.1, 1.0, 1.5, and 2.5.}
\label{5}    
\end{center}
\end{figure}

Few comments regarding the observation of the accelerated inertial regime are in order: (i) We emphasize that the incorporation of hydrodynamics is essential to observe the inertial growth. This is demonstrated in Fig.~\ref{5}(a) which shows $\ell_s(t)$ vs. $t$ for the NHT, DPD and LT. Clearly the stochastic LT does not give rise to inertial growth, rather exhibits the Lifshitz-Slyozov law: $\ell_s(t) \sim t^{1/3}$. On the other hand, the hydrodynamics preserving NHT and DPD lead to the $t^{2/3}$ behaviour. (ii) Next, from Fig.~\ref{5}(b), it is clear that the inertial growth regime stretches over longer time windows in larger system sizes. (iii) The novel accelerated inertial growth and triggered magnetic order is characteristic of the spinodal region where coarsening morphologies are bi-continuous. This evident from Fig.~\ref{5}(c) which shows $\ell_s(t)$. vs. $t$ for $\rho=0.05$ which falls in the nucleation region. The dashed line with slope $t^{1/3}$ is a guide to the eye. The inset shows the snapshot of the system at $t=60$ that is obtained from a typical homogeneous initial state such as that in Fig.~\ref{1}(a). (iv) Finally, in Fig.~\ref{5}(d) we show the effect of the strength of the dipole moment on the inertial growth for several values of $\mu$. The dashed line with slope 2/3 is a guide to the eye. The data for $\mu \gtrsim 1.0$ is well represented by the $t^{2/3}$ law. The dipole-dipole interactions are a crucial ingredient for accelerated inertial growth and a quantitative explanation is provided in the forthcoming discussion. 

\begin{figure}
    \begin{center}
    \includegraphics[width=1\linewidth]{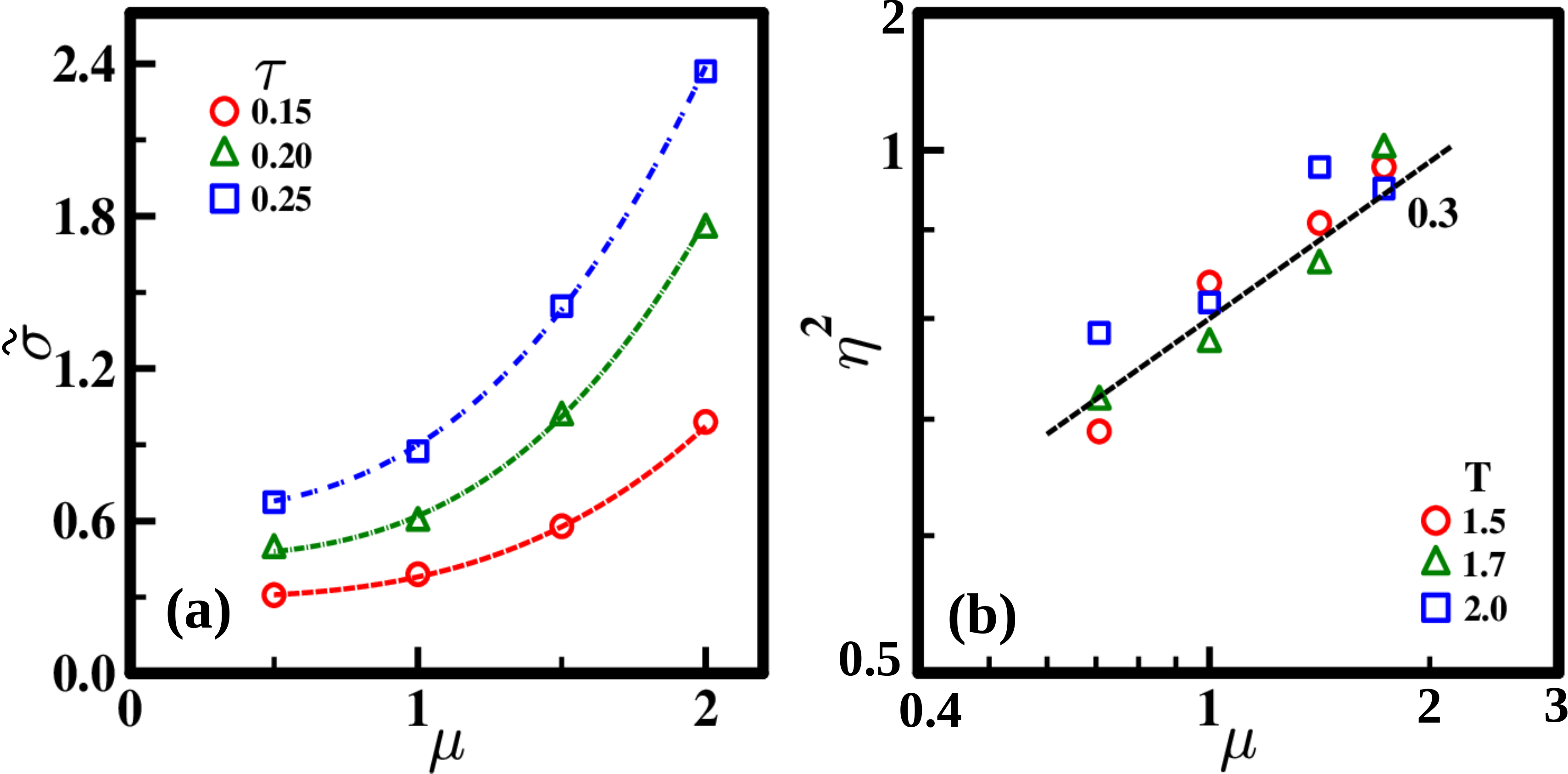}
    \caption{(a) Variation of $\tilde{\sigma}$ vs. $\mu$ for specified values of the reduced temperature $\tau$ for $\rho \lesssim \rho_c$ read from \cite{groh1999}. Dashed curves are fits to $\tilde{\sigma} \approx \tilde{\sigma}_0 + a\mu^{3}$. (b) Variation of $\eta$ vs. $\mu$ for $\rho$ = 0.6 from \cite{eggebrecht1987}. Dashed line with slope 0.3 is a guide to the eye. All data are in LJ units.}
\label{6}    
\end{center}
\end{figure}

\section{Discussion and Summary}
What leads to accelerated inertial growth in the SM fluid? The primary factors governing the accessibility of the hydrodynamic regimes in an incompressible fluid are shear viscosity $\eta$ and surface tension $\tilde{\sigma}$ \cite{puri1992}. For instance, the cross-over from viscous to inertial regime is estimated at $\ell_{s}^*=\eta^2/\rho \tilde{\sigma}$ [and $t_{s}^*=\left(\ell_{s}^*\right)^{3/2}$]. Many groups have studied the variation of $\tilde{\sigma}$ and $\eta$ with respect to $\mu$, $\rho$ and $T$ \cite{groh1999,abbas1998,eggebrecht1987,nagy2020}. Fig.~\ref{6}(a) shows the variation of $\tilde{\sigma}$ with respect to $\mu$ for specified values of the reduced temperature $\tau$ = $(1 - T /T_c )$ from \cite{groh1999}. These data obtained using density functional theory (DFT), have been reconfirmed by MD, DFT and hybrid MD-DFT simulations in \cite{abbas1998}. They are well-represented by $\tilde{\sigma} \approx \tilde{\sigma}_0 + a\mu^{3}$ where $\tilde{\sigma}_0$ is the surface tension of the LJ fluid \cite{groh1999}. Fig.~\ref{6}(b) shows the variation of $\eta$ with respect to $\mu$ for $\rho$ = 0.6 and specified values of temperature $T$, has been obtained in \cite{eggebrecht1987} via MD simulations. The dashed line with slope 0.3 suggests that $\eta^2 \sim \mu^{0.3}$. (As seen from the data in \cite{groh1999,abbas1998}, dipolar effects are negligible for $\mu\lesssim0.5$ and the behaviour is more like the LJ fluid.) It is therefore reasonable to assume that  $\ell_{s}^*\sim\mu^{-2.7}$ and $t_{s}^*\sim\mu^{-4.1}$. So $\ell_{s}^*\simeq 6.498$ and $t_{s}^* \simeq 17.148$ for $\mu= 0.5$, while $\ell_{s}^*\simeq 0.335$ and $t_{s}^*\simeq 0.190$ for $\mu=1.5$. The decrease by 94.8\% in $\ell_{s}^*$ and 98.9\% in $t_{s}^*$ is indeed dramatic! We have thus identified a fluid with overwhelming inertial hydrodynamics from the outset. We trace it's origin to increased surface tension due to dipole-dipole interactions.

To conclude, dipolar fluids exhibit a GL phase transition, anisotropic structures and magnetic order even in the absence of external fields. Common examples of dipolar fluids are low-molecular-weight liquid crystals, ferrofluids, and polymers. They are interesting for theoretical studies, suitable for scientific applications, and hold technological promise. The SM model, incorporating the LJ potential and long-range dipolar interactions, captures the basic features of these fluids. We quench this system in the coexistence region, and study the non-equilibrium phenomenon of coarsening using MD simulations. The Ewald summation technique has been used to accurately evaluate the dipolar interactions. The fluid inertia overpowers the capillary and viscous forces, and the liquid phase grows as $\ell_s(t)\sim t^{2/3}$ in the spinodal region. The predicted inertial growth regime has never been detected in ($d=3$) MD simulations, and this makes our observations significant.  We also see the development of magnetic order in the condensed liquid which is consistent with the $\ell_M(t)\sim t$ prediction in dipolar systems. These observations can trigger inquiries in fundamental science and technological applications. For example, the manipulation of the spatial and magnetic order in the homogeneous liquid phase can be useful in switching applications. Anisotropic shapes with anisotropic interactions emerging in this microscopic framework can lead to a new class of mesoscopic magnetic colloids. We hope that our work sows the seeds for such investigations.  

\vspace{0.25cm}
\noindent {\bf Author Contributions}: VB formulated the problem. AKS performed the numerical simulations. AKS and VB did the analysis and wrote the paper.\\
\noindent {\bf Conflicts of interest}: There are no conflicts of interests to declare.\\
\noindent {\bf Acknowledgements}: We thank Sanjay Puri, Subir Das, Gaurav Prakash Shrivastava and Arunkumar Bhupathy for valuable discussions. The HPC facility at IIT Delhi is gratefully acknowledged for computational resources. VB acknowledges SERB (India) for CORE and MATRICS grants.





\bibliographystyle{apsrev4-1}
\bibliography{biblio.bib}

\end{document}